\documentclass{iau} 
\usepackage{graphicx}

\title[JD 11.~~Tomography of the red supergiant star $\mu$ Cep] %% give here short title %%
{Tomography of the red supergiant star $\mu$ Cep}

\author[K. Kravchenko et al.]   %% give here short author list %%
{K. Kravchenko$^{1,2}$, A. Chiavassa$^3$, S. Van Eck$^2$, A. Jorissen$^2$, T.~Merle$^2$, B. Freytag$^4$
 }

\affiliation{$^1$European Southern Observatory, Karl-Schwarzschild-Str. 2, 85748 \\ Garching bei M{\"u}nchen, Germany\\email: {\tt kateryna.kravchenko@eso.org} \\[\affilskip]
		   $^2$ Institut d'Astronomie et d'Astrophysique, Universit\'e Libre de Bruxelles, \\ CP226, Boulevard du Triomphe, 1050 Bruxelles, Belgium\\[\affilskip]
           $^3$ Universit\'e C\^ote d'Azur, Observatoire de la C\^ote d'Azur, CNRS, Lagrange, \\ CS 34229, 06304 Nice Cedex 4, France\\[\affilskip]
		   $^4$ Department of Physics and Astronomy at Uppsala University, \\ Regementsv{\"a}gen 1, Box 516, 75120 Uppsala, Sweden	       }
\pubyear{2018}
\volume{xxx}  %% insert here IAU Symposium No.
\jname{Why Galaxies Care About AGB Stars: A Continuing Challenge through Cosmic Time}
\editors{}
\begin{document}

\maketitle

\begin{abstract}

A tomographic method, aiming at probing velocity fields at depth in stellar atmospheres, is applied to the red supergiant star $\mu$ Cep and to snapshots of 3D radiative-hydrodynamics simulation in order to constrain atmospheric motions and relate them to photometric variability.

\end{abstract}

\firstsection % if your document starts with a section,
\section{Introduction}

Red supergiant (RSG) stars show irregular photometric variations characterized by two main periods (\cite[Kiss \etal\ 2006]{Kiss_etal2006}): a short one (few hundred days) and a longer one (few thousand days). We focus on the short photometric period which was proposed to be due to either stellar pulsations or convection. We aim at constraining the atmospheric motions in the RSG star $\mu$ Cep and at relating them to photometic variability using a tomographic method (\cite[Kravchenko \etal\ 2018]{Kravchenko_etal2018}).

{\underline{\textbf{Tomographic method}}} aims at recovering the line-of-sight velocity field as a function of the optical depth in the atmosphere. The method is based on sorting spectral lines according to their formation depth provided by the contribution function (CF) maximum at each wavelength. This allows to split the atmosphere into different layers and construct masks which contain lines forming in the corresponding ranges of optical depths. The cross-correlation of masks with stellar spectra provides the velocity in the different atmospheric layers.

\section{Application to $\mu$ Cep and 3D radiative-hydrodynamics (RHD) simulations}

$\mu$ Cep was observed with the HERMES spectrograph (\cite[Raskin \etal\ 2011]{Raskin_etal2011}, R $\sim$ 86000).  95 spectra were obtained between April 2011 and January 2018, corresponding to a time span of about 7 years. A set of 5 tomographic masks was constructed from a 1D MARCS (\cite[Gustafsson \etal\ 2008]{Gust_etal2008}) model atmosphere with $T_{\rm eff} = 3400$~K and $\log g = -0.4$. The masks were cross-correlated with $\mu$ Cep spectra in order to provide RV in each atmospheric layer.
%corresponding RVs were derived. 

\begin{figure}[h]
%% \vspace*{-2.0 cm}
\begin{center}
 \includegraphics[width=6.2cm]{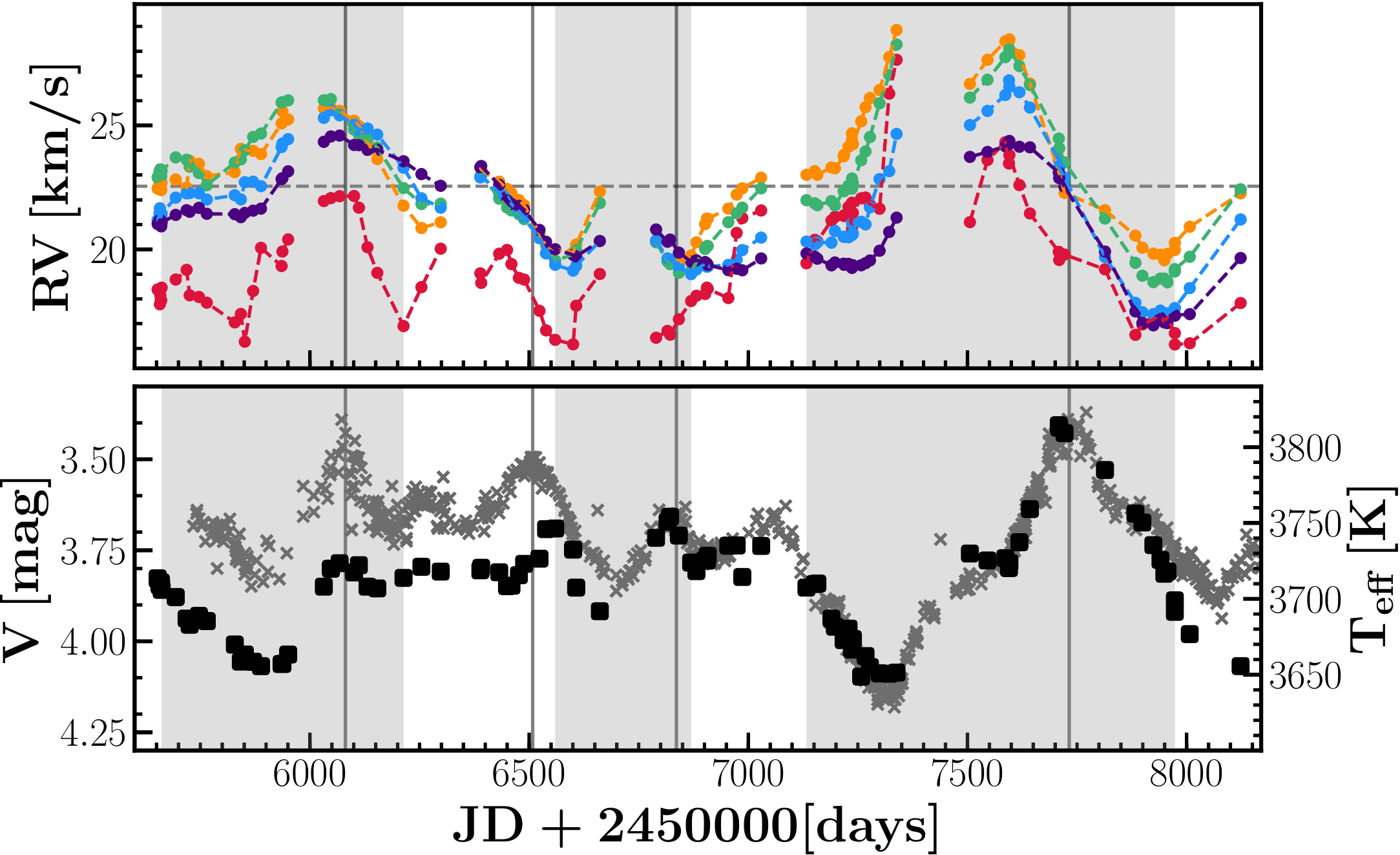} 
  \includegraphics[width=5.7cm]{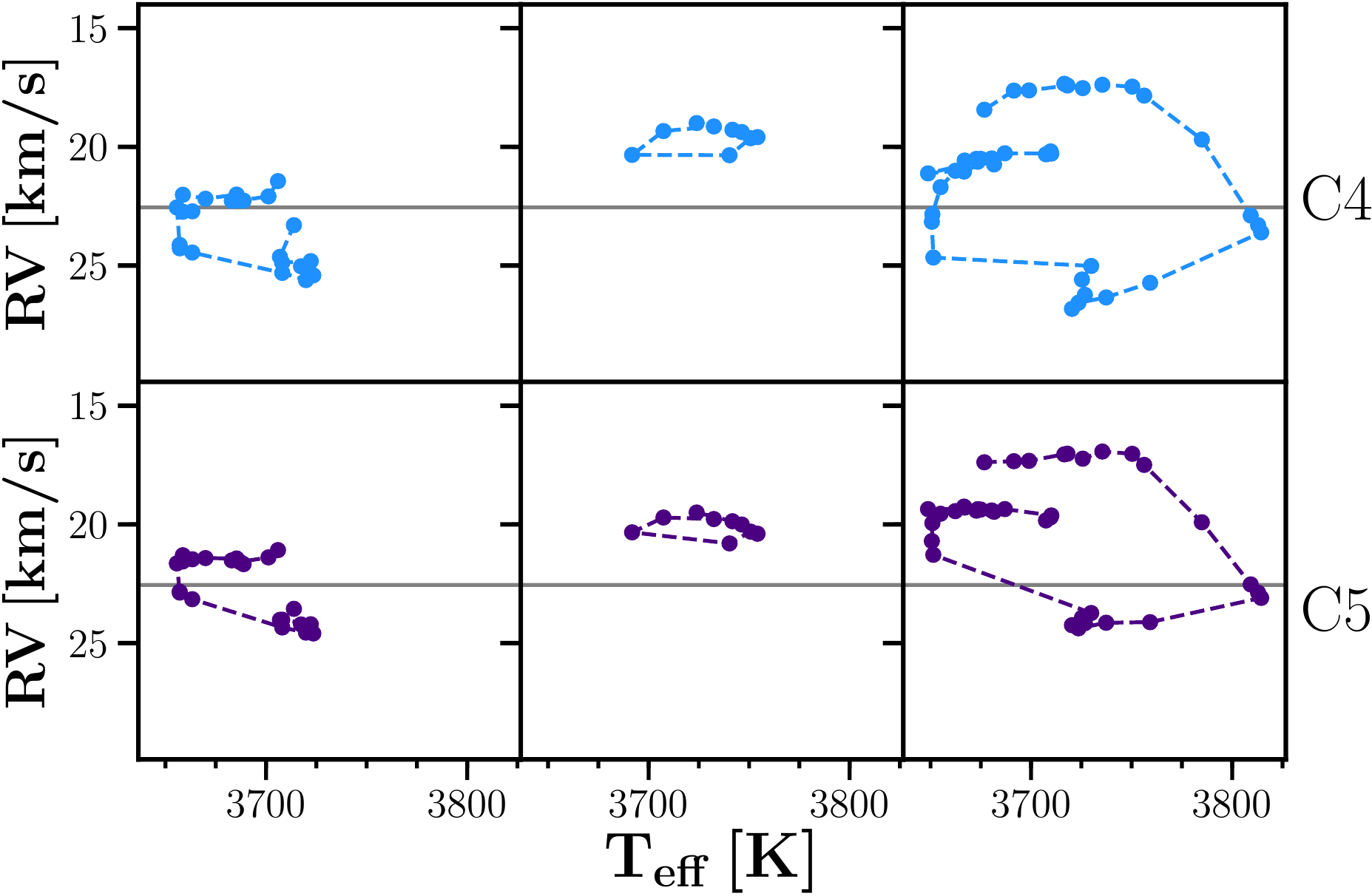} 
%% \vspace*{-1.0 cm}
 \caption{\textit{Left panel:} The RVs in different masks compared to the AAVSO visual light curve (crosses) and $\rm T_{eff}$ (squares). Vertical lines indicate times of a maximum light. Shaded areas define pseudo light cycles which correspond to hysteresis loops. \textit{Right panel:} Hysteresis loops for masks C4 and C5 (rows) corresponding to three pseudo-cycles (columns).
 }
   \label{fig1}
\end{center}
\end{figure}

The RVs in masks C1-C5 are compared with the AAVSO light curve and the effective temperature ($\rm T_{eff}$) in Fig.\,\ref{fig1}. A phase shift of about 100 days is observed between RV and V magnitude (and $\rm T_{eff}$) variations. This phase lag results in hysteresis loops in the $\rm T_{eff}-RV$ plane (Fig.\,\ref{fig1}) with timescales similar to the photometric ones. Similar hysteresis loops were observed by \cite[Gray (2008)]{Gray_2008} for Betelgeuse and interpreted as the turn-over of the material through a large convective cell.

\begin{figure}[h]
%% \vspace*{-2.0 cm}
\begin{center}
 \includegraphics[width=5.8cm]{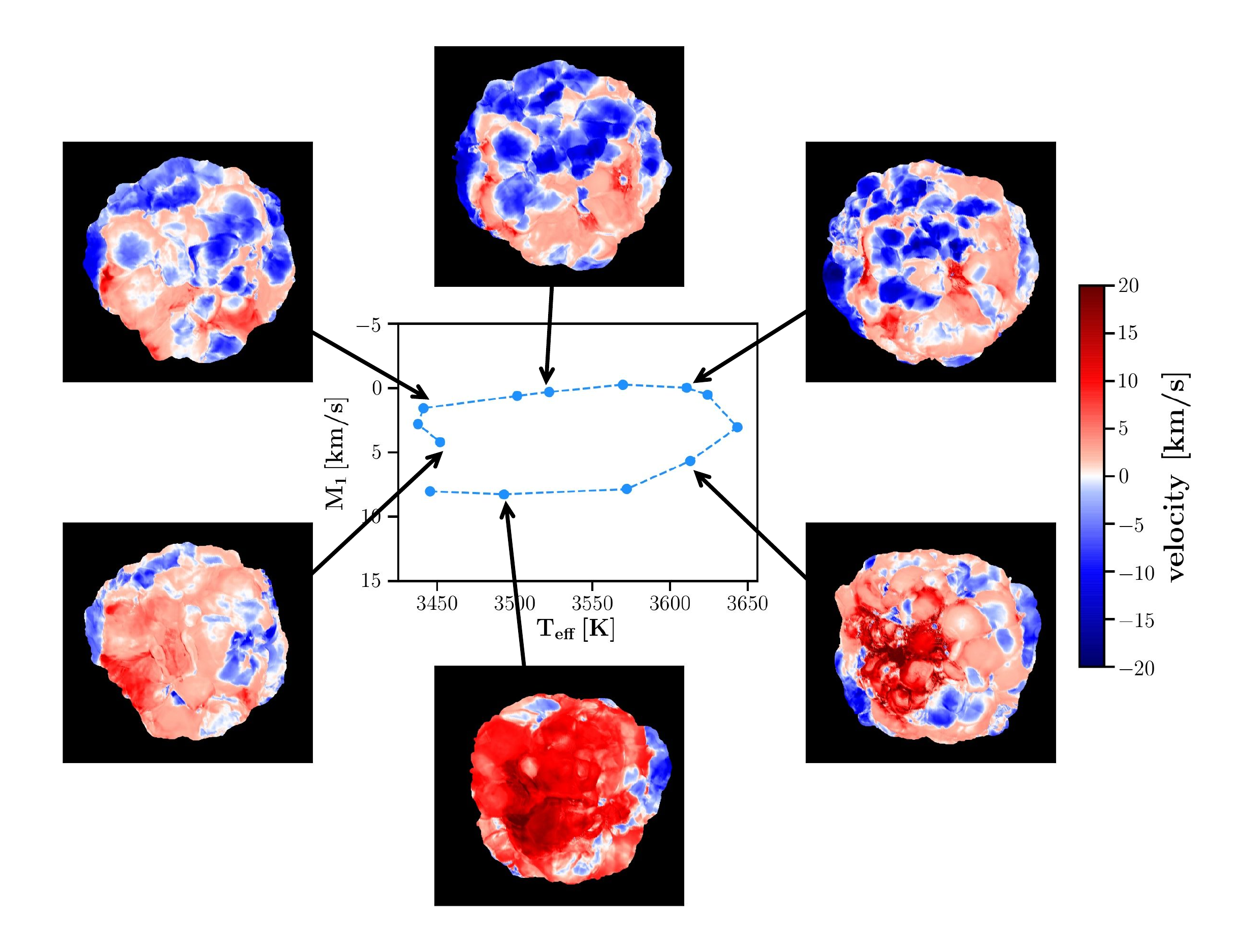} 
 \includegraphics[width=5.8cm]{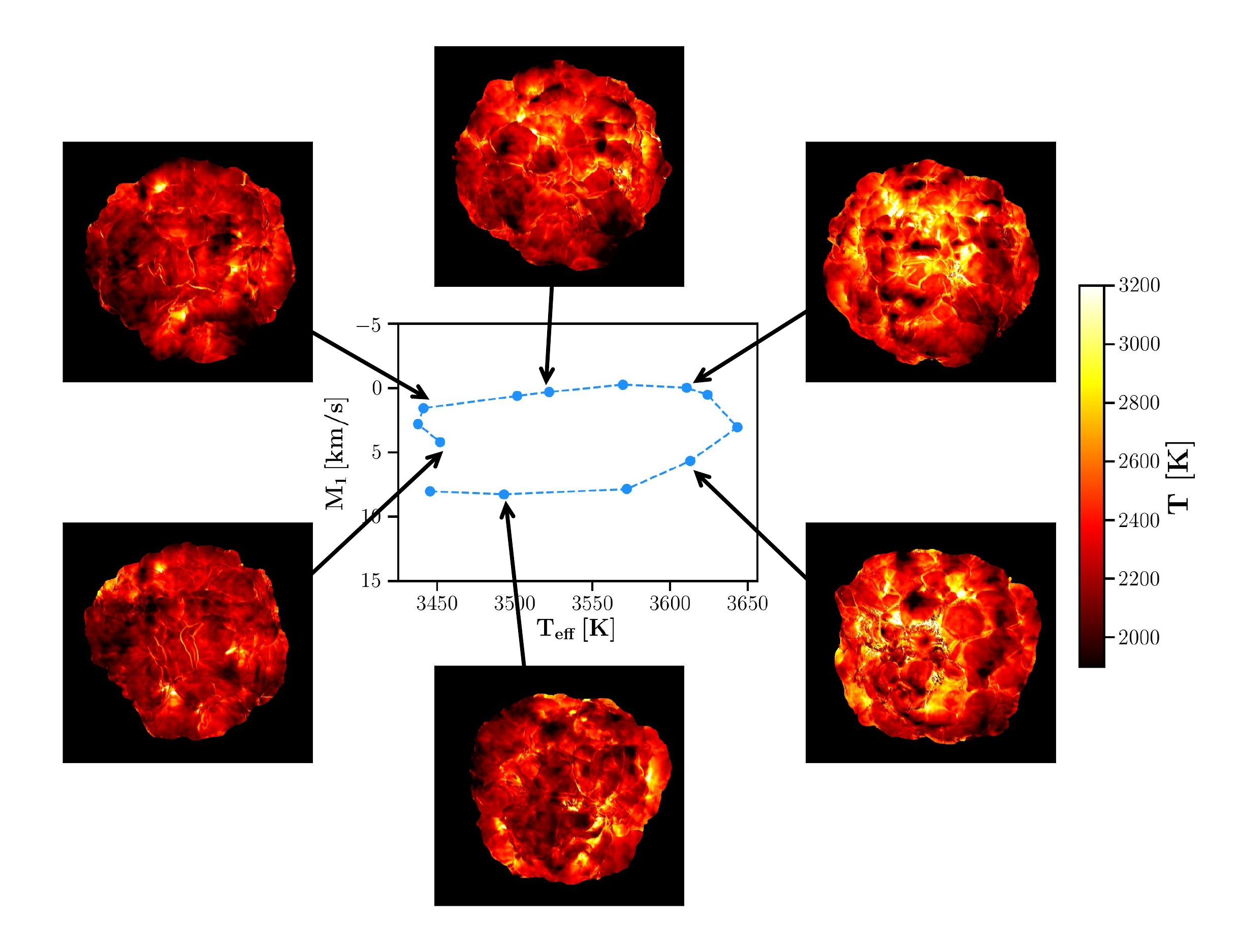}
%% \vspace*{-1.0 cm}
 \caption{Velocity (left) and temperature (right) maps weighted by the CF of a line from mask C4 for snapshots along the hysteresis loop. }
   \label{fig2}
\end{center}
\end{figure}

The tomographic method was applied to snapshots from the 3D RHD simulation of a RSG star performed with the CO5BOLD code (\cite[Freytag \etal\ 2012]{Freytag_etal2012}). Hysteresis loops in the $\rm T_{eff}-RV$ plane were detected and show timescales similar to the photometric ones. Velocity and temperature maps for snapshots along the hysteresis loop of mask C4 were weighted by the CF of a spectral line from the same mask. They are shown in Fig.\,\ref{fig2} and reveal convective motions in the atmosphere. The consistency between the hysteresis loop timescales measured in $\mu$ Cep and in the 3D simulation indicates that convection might account for the short-period photometric variations in $\mu$ Cep.

\end{document}